\documentclass[a4paper]{jpconf}
\usepackage{graphicx}
\usepackage{euscript,graphicx,amssymb,subfigure}

\usepackage{epsfig}
\usepackage{amsfonts}
\usepackage{amsmath}
\usepackage{amssymb}
\usepackage{euscript}
\usepackage{color}
\usepackage{subfigure}

\def\lsim{\mathrel{\rlap{\lower3.5pt\hbox{\hskip0.5pt$\sim$}}
    \raise0.5pt\hbox{$<$}}}                
\def\gsim{~\rlap{$>$}{\lower 1.0ex\hbox{$\sim$}}}

\def\Om{\mbox{$\Omega_{\rm m0}$}}
\def\weff{\mbox{$w_{\rm eff}$}}
\def\peff{\mbox{$p_{\rm eff}$}}

\begin{document}


\title{Constraining decaying dark energy density models with the CMB temperature-redshift relation}

\author{Philippe Jetzer and Crescenzo Tortora}

\address{Universit$\ddot{a}$t Z$\rm\ddot{u}$rich, Institut
f$\rm\ddot{u}$r Theoretische Physik, Winterthurerstrasse 190,
CH-8057, Z$\ddot{u}$rich, Switzerland}


\begin{abstract}
We discuss the thermodynamic and dynamical properties of a variable dark
energy model with density scaling as $\rho_{x} \propto (1+z)^{m}$, z being
the redshift. These models lead to the
creation/disruption of matter and radiation, which affect the
cosmic evolution of both matter and radiation components in the
Universe. In particular, we have studied the
temperature-redshift relation of radiation, which has been
constrained using a recent collection of cosmic microwave
background (CMB) temperature measurements up to $z \sim 3$.
We find that, within the uncertainties, the
model is indistinguishable from a cosmological constant which does
not exchange any particles with other components.
Future observations, in particular measurements of
CMB temperature at large redshift, will allow to give firmer bounds
on the effective equation of state parameter \weff\ for such types of dark
energy models.
\end{abstract}

\section{Introduction}

The current standard cosmology model relies on the existence of
two unknown dark components, the so called ``dark matter'' (DM)
and ``dark energy'' (DE), which amount to $\sim 25\%$ and $\sim
70\%$ of the total energy budget in the Universe, respectively.
According to several observations, the Universe is spatially flat
and in an accelerated phase of its expansion \cite{perl, reiss,
debernardis, spe07, caldwell}. DE, described as a cosmological
constant $\Lambda$ in its simplest form, is modelled by a fluid
with a negative pressure, which is a fundamental ingredient to
explain the actual accelerated expansion of the Universe.

Several models have been proposed to explain DE \cite{PR88, RP88, SS00,
Caldwell02, Pad03, PR03, D+05, CTTC06}. An alternative
consists to consider a phenomenological variable DE density with
continuous creation/disruption of photons \cite{lima+96, LA99,
lima+00, puy, Jetzer+11, Jetzer} or matter \cite{CW90, ma}. The DE might
decay/grow slowly in the course of the cosmic evolution and thus
provide the source/sink term for matter and radiation. Different
such models have been discussed and strong constraints come from
very accurate measurements of the cosmic microwave background
(CMB) radiation and other typical cosmological probe.

CMB radiation is the best evidence for an expanding Universe
starting from an initial high density state. Within the
Friedmann-Robertson-Walker (FRW) models of the Universe the
radiation, after decoupling, expands adiabatically and scales as
$(1+z)$, $z$ being the redshift \cite{Weinberg72}. If we assume
that each component is not conserved, contrarily to the standard
scenario, then depending on the decay mechanism of the DE, the
created photons could lead to distortions in the Planck spectrum
of the CMB, and change the evolution of its temperature. The
chance to appreciate the deviation from the standard temperature
evolution is given by the increasingly number of recent works
collecting observations of CMB temperature both at low
\cite{battistelli, luz09} and higher redshifts
\cite{Noterdaeme+11}.

Following the theoretical lines of \cite{lima+96, LA99, lima+00,
puy}, in Jetzer et al. \cite{Jetzer+11, Jetzer} we have discussed a
variable DE model $\Lambda (z) \propto (1+z)^{m}$ decaying into
photons and DM particles. In particular we studied thermodynamical
aspects in the case of a continuous photon creation, which implies
a modified temperature redshift relation for the CMB. We have
tested the predicted temperature evolution of radiation with some
recent data on the CMB at higher redshift from both
Sunyaev-Zel'dovich (SZ) effect and high-redshift QSO absorption
lines as well as with an updated collection of
data from
different kind of observations, like distance moduli of Supernovae
Ia, observations of the CMB anisotropy and the large-scale
structure, together with observational Hubble parameter
estimations \cite{Jetzer}. A similar approach, however without
considering photon creation and thus a modification in the CMB temperature
evolution, has been discussed by Ma \cite{ma}.

\section{Theoretical framework}\label{sec:theory}

We assume a cosmological framework  based on the
usual Robertson-Walker (RW) metric element \cite{Weinberg72}
and that the Universe contains
three different components: a) a matter (both baryons and DM)
fluid, with equation of state $p_{m}=0$ (since $p_{m} <<
\rho_{m}$), b) a generalised fluid with pressure $p_{\gamma} =
(\gamma - 1) \rho_{\gamma}$, where $\gamma$ is a free parameter,
which is set to $4/3$ for a properly said radiation fluid, and c)
a DE, $x$ component, with pressure $p_x$ and density $\rho_x$. The
equation of state for the $x$ component could assume a very general
expression, but we limit ourselves to consider the simple linear relation
$p_{x} = w_{x} \rho_{x}$. We will set any 'bare' cosmological
constant $\Lambda_0$ equal to 0 \cite{Jetzer+11}. With these
components we get for the Einstein field equations
\cite{Weinberg72}

\begin{equation}
8\pi G \rho_{tot} = 3\frac{\dot R^2}{R^2}+3 \frac{k}{R^2}~,
\label{eq:Fried_1}
\end{equation}
\begin{equation}
8\pi G p_{tot} = -2\frac{\ddot R}{R}- \frac{\dot
R^2}{R^2}-\frac{k}{R^2}~, \label{eq:Fried_2}
\end{equation}
where $p_{tot} = p_{\gamma} +p_x$ and $\rho_{tot} = \rho_{m} +
\rho_{\gamma} + \rho_x$ are the total pressure and density, $R$ is
the scale factor, $k=0, \, \pm 1$ is the curvature parameter and a
dot means time derivative.
Furthermore, we will assume that there is no curvature,
thus $k=0$ \cite{debernardis}.

In the following we will adopt $w_x=-1$, however since we are
assuming that the vacuum decays into radiation and massive
particles, the effective equation of state $w_{eff}$, which is the
measured quantity, can differ from $-1$.

Following \cite{Jetzer+11, Jetzer} (but see also
\cite{lima+96, lima+00}) we set here the energy conservation
equation for the different fluids. The fluid as whole, verifies
the Bianchi identity, which means that energy and momentum are
locally conserved: $\nabla_{\mu} T^{\mu \nu} =
0$, with $T^{\mu \nu}$ the stress-energy tensor
\begin{equation}
T^{\mu\nu} = (\rho_{tot}+p_{tot})u^{\mu}u^{\nu}-p_{tot}g^{\mu\nu}
\end{equation}
and $u^{\mu}$ being the four velocity. After easy calculations,
this conservation law reads
\begin{equation}
\dot\rho_{tot} +3 (\rho_{tot}+p_{tot}) H = 0  ~,\label{eq:cons_ALL}
\end{equation}
where $H = \dot R / R$ is the Hubble parameter. In the standard
approach, each component is conserved, thus Eq.
(\ref{eq:cons_ALL}) holds for all the fluid components, but here
we will suppose that each component will exchange energy with each
other. In particular, for matter, radiation and DE we impose the
following relations (see also \cite{mubasher})
\begin{equation}
\dot\rho_{m} +3 \rho_{m} H = (1-\epsilon) ~ C_{x}
~,\label{eq:cons_mat}
\end{equation}
\begin{equation}
\dot\rho_{\gamma} +3 \gamma \rho_{\gamma} H = \epsilon ~ C_{x}
~,\label{eq:cons_gamma}
\end{equation}
\begin{equation}
\dot\rho_{x} +3(p_{x} + \rho_{x})H = - C_{x} ~,\label{eq:cons_x}
\end{equation}
where we assume that both the matter and the radiation fluids
exchange energy with the DE as parameterized by $C_{x}$ and
$\epsilon$. In particular, $C_{x}$ depends on the DE and, indeed,
acts as a source/sink term for the fluids energy. Evidently, if no
interaction between the different components exists, then $C_x$ is
null and the standard picture is recovered. Moreover, if $\epsilon
= 0$ ($\epsilon = 1$), then the DE exchanges all the energy with
matter (radiation). $C_x$ can describe different physical
situations such as, for instance, a thermogravitational quantum
creation theory \cite{LA99} or a quintessence scalar field
cosmology \cite{RP88}. As we will discuss later on, $\epsilon$ has
to be very small (i.e., $\epsilon \ll 1$), otherwise the radiation
density would become much too big, contrary to present values. As
an order of magnitude estimate we expect $\epsilon \simeq
\frac{p_{\gamma} + \rho_{\gamma}}{\rho_{m}}$, for which indeed
$\epsilon \ll 1$, since $p_{\gamma} + \rho_{\gamma} \ll \rho_{m}$.

Adopting as mentioned above the relation $p_x=-\rho_x$ and defining
$\rho_x=\Lambda(t)/(8\pi G)$, from Eq.
(\ref{eq:cons_x}) we obtain
\begin{equation}
C_x= -\frac{\dot\Lambda(t)}{8\pi G}~. \label{eq:Cx}
\end{equation}
We assume a power law model for the $\Lambda(t)$ function,
$\Lambda(t) = B (R / R_{0})^{-m}$, or equivalently in terms of
redshift, $\Lambda(z) = B (1+z)^{m}$, where $B$ is a constant. The
value of this constant is $B=3 H_{0}^{2} (1- \Om)$, which can be
found using Eq. (\ref{eq:Fried_1}) at the present epoch, assuming
that today $\rho_{\gamma} \ll \rho_{m}, \rho_{x}$. Thus, the
density evolution for the $x$ component is given by
\begin{equation}
\rho_{x}/\rho_{crit}  = (1 - \Om) (1+z)^{m},
\end{equation}
where we have defined as $\rho_{crit} = \frac{3 H_{0}}{8 \pi G}$
the present critical density of the Universe. If $m$ is positive,
then the DE slowly decreases as a function of the cosmic time,
whereas if $m$ is negative the inverse process happens.

From Eqs. (\ref{eq:cons_mat}) and (\ref{eq:cons_gamma}) we derive
the evolution laws for matter and radiation,
\begin{align}\label{eq:rho_m}
\rho_{m}/\rho_{crit} & = \Om (1+z)^{3}
\\ \nonumber
& \quad - (1-\epsilon) \frac{m (1- \Om)}{m - 3} [(1+z)^{m} -
(1+z)^{3}]~,
\end{align}
\begin{align}\label{eq:rho_gamma}
\rho_{\gamma}/\rho_{crit} & = \Omega_{\gamma 0} (1+z)^{3 \gamma}
\\ \nonumber
& \quad - \epsilon \frac{m (1- \Om)}{m - 3 \gamma} [(1+z)^{m} -
(1+z)^{3 \gamma}]~,
\end{align}
where \Om\ and $\Omega_{\gamma 0}$ are the matter and radiation
energy densities at $z=0$, respectively.

Since we are interested in the evolution of the radiation
temperature, it is useful to discuss the extreme case when only
photons enter in the process (i.e. $\epsilon = 1$). Then
for the matter density in Eq. (\ref{eq:rho_m}) the usual evolution
$\propto (1+z)^{3}$ holds, while the radiation besides the usual
term $\Omega_{\gamma 0} (1+z)^{3 \gamma}$ has also a perturbative term
depending on $m$. Since today \footnote{The present radiation density
is the only cosmological parameter accurately measured. The
radiation density is dominated by the energy in the cosmic
microwave background (CMB), and the COBE satellite FIRAS
experiment determined its temperature to be $T = 2.725 \pm 0.001
\, K$ \cite{Mather+99}, corresponding to $\Omega_{\gamma 0} \sim  5
\times 10^{-5}$.} $\Omega_{\gamma 0}   \sim 5 \times 10^{-5}$
it turns out that $m$ has to be extremely small $\lsim 10^{-4}$.
Therefore, unless $m$ is extremely small or vanishing, DE
has to decay mainly in matter with possibly some photons as well.
Thus the condition $\epsilon \ll 1$ has to hold.

\subsection{Hubble and deceleration parameter}

Due to the very small value of $\Omega_{\gamma 0}$ it follows that
the evolution of the Universe is essentially driven by the DE and
DM, therefore, from Eq. (\ref{eq:Fried_1}) the following law for
the Hubble parameter holds
\begin{align}\label{eq:H}
H(z) & \simeq  \frac{8 \pi G}{3} (\rho_{m} + \rho_{x})
\\ \nonumber
& =  H_0 \left[\frac{3 (1-\Om)}{3 - m} (1+z)^{m} + \frac{(3\Om -
m) }{3-m} (1+z)^{3} \right]^{1/2}~,
\end{align}
which is obviously the same expression found in Ma \cite{ma}.

Recasting Eq. (\ref{eq:cons_x}), it is possible to write
\begin{equation}
\dot\rho_{x} +3H(p_{x} + \rho_{x} +\frac{C_{x}}{3 H}) = 0
~,\label{eq:cons_x_2}
\end{equation}
which shows that the term $C_x$ contributes to an effective
pressure
\begin{equation}
\peff = p_{x} + \frac{C_{x}}{3 H} = -\rho_{x} +\frac{C_{x}}{3 H}~.
\end{equation}
Therefore, we get an equivalent
effective DE equation of state $\weff$ \cite{ma}
\begin{equation}
\weff = \frac{\peff}{\rho_x} = \frac{m}{3}-1 .
\end{equation}
If $m>0$ then we have $\weff
> -1$, i.e. our model is quintessence-like \cite{PR88,RP88,D+05},
while we have a phantom-like \cite{Caldwell02} model when $m$ is
negative and $\weff<-1$. Another interesting quantity is the
deceleration parameter, which can be written as
\begin{align}
q(z) & = - \frac{\ddot{R}R}{\dot{R}^{2}}= \\ \nonumber &
\frac{(1+z)^{3}(m-3\Om)+3(m-2)(1+z)^{m}(\Om-1)}{2(1+z)^{3}(m-3\Om)+6(1+z)^{m}(\Om-1)}~.
\end{align}
Imposing that $q(z)=0$, we can determine the {\it transition
redshift}, i.e. the redshift at which the Universe changed from a
deceleration to an acceleration phase, which is given by
\begin{equation}
z_{T}=\bigg ( \frac{3(2-m)(1-\Om)}{3\Om-m}
\bigg)^{\frac{1}{3-m}}-1~.
\end{equation}
From this result we see that the larger $m$ is, the earlier the
Universe changes from deceleration to acceleration.

\subsection{Thermodynamical aspects and CMB temperature evolution}\label{sec:T_relation}

We follow here the approach outlined in Lima \cite{lima+96}, where
he defines a current as $N^{\alpha}=n u^{\alpha}$ with $n$ being
the particle number density of the photons or of the DM particles.
Indeed, there is a current for each of these components. Due to
the decaying vacuum the current satisfies the following balance
equation (one for each component)
\begin{equation}
\dot n_i + 3n_iH= \psi_i~, \label{nn}
\end{equation}
with $i=\gamma$ or $DM$ ($\gamma$ for the photons and $DM$ for the
dark matter) and $\psi_i$ is the corresponding particle source.
For decaying vacuum models $\psi_{\gamma}+\psi_{DM}$ is positive
and related to the rate of change of $\rho_x$. We can also define
an entropy current of the form
\begin{equation}
S^{\alpha}=\sum_i n_i \sigma_i u^{\alpha}_i~,
\end{equation}
where $\sigma_i$ is the specific entropy per particle (photons, DM
and in principle also DE). If the DE $\rho_x$ is constant the
above entropy current is conserved. The existence of a non
equilibrium decay process of the vacuum implies
$S^{\alpha}_{;\alpha} \geq 0$, thus an increase of the entropy as
a consequence of the second law of thermodynamics. In principle
the second law should be applied to the system as a whole, thus
including the vacuum component \cite{mubasher}. Assuming that the
vacuum is like a condensate with zero chemical potential
$\mu_{vac}$ it follows from Euler's relation
\begin{equation}
\mu_{vac}= \frac{\rho_x+p_x}{n}-T \sigma_{vac}~,
\end{equation}
provided $w_x=-1$, that $\sigma_{vac}=0$ and thus
its contribution to the entropy current vanishes.
Given our assumptions the vacuum plays the role
of a condensate carrying no entropy.

For instance, for quintessence models in the limit where the
scalar field does not depend on time, and thus its time derivative
vanishes, one gets $w_x=-1$. In the later stages of the Universe
the time dependence is possibly very weak so that $w_x=-1$ holds
up to small corrections.

The equation for the particle number density of radiation
component is given by Eq. (\ref{nn})
with $i=\gamma$.
Using Gibbs law and well-known
thermodynamic identities, following the derivation given in the
paper by Lima et al. \cite{lima+00}, one gets (see also
\cite{lima+96, Jetzer+11})
\begin{equation}
\frac{\dot T}{T} = \left(\frac{\partial p_{\gamma}}{\partial
\rho_{\gamma}}\right)_n \frac{\dot n_{\gamma}}{n_{\gamma}} -
\frac{\psi_{\gamma}}{n_{\gamma} T \left(\frac{\partial
\rho_{\gamma}}{\partial T}\right)_n}
\left[p_{\gamma}+\rho_{\gamma}- \frac{n_{\gamma} \epsilon \,
C_x}{\psi_{\gamma}}\right]~. \label{1}
\end{equation}

To get a black-body spectrum the second term in brackets in Eq.
(\ref{1}) has to vanish, thus

\begin{equation}
\epsilon \, C_x = \frac{\psi_{\gamma}}{n_{\gamma}} \left[
p_{\gamma}+\rho_{\gamma} \right]~. \label{3}
\end{equation}

Thus, Eq. (\ref{1}) becomes

\begin{equation}
\frac{\dot T}{T} = \left(\frac{\partial p_{\gamma}}{\partial
\rho{\gamma}}\right)_n \frac{\dot n_{\gamma}}{n_{\gamma}}~.
\label{5}
\end{equation}

With $\left(\frac{\partial p_{\gamma}}{\partial
\rho_{\gamma}}\right)_n =(\gamma -1)$ one obtains

\begin{equation}
\frac{\dot T}{T} = (\gamma -1) \frac{\dot
n_{\gamma}}{n_{\gamma}}~. \label{6}
\end{equation}

Using the equation for the particle number conservation Eq.
(\ref{nn}) into Eq. (\ref{6}) leads to

\begin{equation}
\frac{\dot T}{T} = (\gamma -1)
\left[\frac{\psi_{\gamma}}{n_{\gamma}}-3H\right]~. \label{eq:T_Psi}
\end{equation}

With Eqs. (\ref{3}) and (\ref{eq:T_Psi}) we get

\begin{equation}
\frac{\dot T}{T} = (\gamma -1) \left[-\frac{\epsilon \,
\dot\Lambda}{8\pi G(p_{\gamma}+\rho_{\gamma})}-3H\right]~.
\label{eq:T_Psi_2}
\end{equation}
Now, following the previous discussion on $\epsilon$ and
aiming to be very general, we set $\epsilon =
\frac{\rho_{\gamma}+p_{\gamma}}{\rho_{m}} \tilde{\epsilon}$, where
$\tilde{\epsilon}$ is a new parameter and insert it into Eq.
(\ref{eq:T_Psi_2}). Taking the sum of Eqs. (\ref{eq:Fried_1}) and
(\ref{eq:Fried_2}) we find
\begin{equation}
8\pi G(\rho_{tot}+p_{tot}) \simeq 8\pi G \rho_{m} =2\frac{\dot
R^2}{R^2}-2 \frac{\ddot R}{R} = -2\dot H~. \label{13}
\end{equation}
Finally, we obtain the expression
\begin{equation}
\frac{\dot T}{T} = (\gamma -1) \left[\frac{\dot\Lambda
\tilde{\epsilon}}{2\dot H}-3H\right]~, \label{14}
\end{equation}
which we can integrate

\begin{equation}
\int^{t_0}_{t_1} \frac{\dot T}{T} dt = (\gamma - 1)
\int^{t_0}_{t_1}\left[\frac{\dot\Lambda \tilde{\epsilon}}{2\dot
H}-3H\right] dt~, \label{15}
\end{equation}
where $t_0$ denotes the present time and $t_1$ some far instant in
the past. Indeed, if $\dot\Lambda$ vanishes and $\gamma=4/3$ one
gets the usual dependence $T(t)=\frac{R(t_1) T(t_1)}{R(t)}$ for a
radiation fluid. To carry out the integration of the first term on
the right hand side it is useful to perform a change of variable
from $t$ to $z$ and accordingly $\frac{dt}{dz}=\frac{-1}{H(1+z)}$.
This way we get (with $z_1$ corresponding to the time $t_1$ and
$z_0=0$ corresponding to $t_0$ present time)
\begin{align}\label{eq:T_fin}
ln \frac{T(z=0)}{T(z_1)}+3(\gamma-1)ln\frac{R(z=0)}{R(z_1)}=
\\ \nonumber \frac{(\gamma-1)}{2} \int_0^{z_1}
\frac{\Lambda^{\prime} \tilde{\epsilon}}{H^{\prime} H(1+z)} dz~,
\end{align}
where $^{\prime}$ denotes derivative with respect to $z$.

As next we insert $H(z)$ and its derivative as taken from Eq.
(\ref{eq:H}) into Eq. (\ref{eq:T_fin}) and integrate it, to get
(setting $z_1=z$)

\begin{equation}
T(z) = T_0 \left(\frac{R_0}{R(z)}\right)^{3(\gamma-1)}
exp\left(\frac{B(1-\gamma)\tilde{\epsilon}}{3H_0^2(\Om-1)}A\right)~,
\label{eq:T_fin_2}
\end{equation}
where
\begin{align}\label{ma2}
A = ln((m-3\Om)+m(1+z)^{m-3}(\Om-1))
\\ \nonumber  -ln((m-3)\Om) ]~.
\end{align}

We can also write Eq. (\ref{eq:T_fin_2}) as
\begin{align}\label{ma3}
T(z) & = T_0 (1+z)^{3(\gamma-1)} \\ \nonumber & \times
\left(\frac{(m-3\Om)+m(1+z)^{m-3}(\Om-1)}{(m-3)\Om}\right)^{\tilde{\epsilon}(\gamma-1)}~.
\end{align}
We inserted in the exponent of Eq. (\ref{eq:T_fin_2}) the explicit
form of $B$, thus getting as exponent in the above Eq.
$\tilde{\epsilon} (\gamma-1)$. Hereafter, we will set
$\tilde{\epsilon} = 1$. Clearly $\tilde{\epsilon}$ and $m$ are not
independent, we checked using the temperature redshift data that
if $\tilde{\epsilon}$ is bigger ($\sim 10$ or more), then $m$ has
to be extremely small consistently with what mentioned above
(as it would lead to a too high production of photons in the
DE decay). On the other hand, if $\tilde{\epsilon}$ gets smaller
(e.g., $\sim 0.1)$ $m$ gets bigger ($\sim 0.2$) and accordingly
$w_{eff}$, moreover m would be poorly constrained, since the
uncertainties would then be very high. But from the other data
(without the the temperature ones) there are already stringent
limits on $m$ and thus this way one could get lower limits on
$\tilde{\epsilon}$, under the assumption that DE decays also in
photons.

Notice that for $z=0$ we have $T(0)=T_0$,
whereas for $m=0$ the expression in the parenthesis is equal to 1
and thus $T(z)=T_0(1+z)^{3(\gamma-1)}$, which for the canonical
value of $\gamma=4/3$ reduces to the standard expression.

\section{Results}\label{sec:results}

To test the temperature evolution for the radiation component, we
rely on the CMB temperatures derived from the absorption
lines of high redshift systems and the ones from SZ effect in
clusters of galaxies (we will collectively quote as $T_{CMB}$,
hereafter). At high redshift the CMB temperature is recovered from
the excitation of interstellar atomic or molecular species that
have transition energies in the sub-millimetre range and can be
excited by CMB photons. When the relative population of the
different energy levels are in radiative equilibrium with the CMB
radiation, the excitation temperature of the species equals that
of the black-body radiation at that redshift, providing one of the
best tools for determining the black-body temperature of the CMB
in the distant Universe \cite{ge97, sri00, mol02, puy93, gal98,
sta98, cui05, sri08}. We used
a sample of 9 QSO absorption measurements \cite{Jetzer,Noterdaeme+11}.
In summary we have 4 data points from the
analysis of the fine structure of atomic carbon (AC) and 5
measurements based on the rotational excitation of CO molecules
(CO) \cite{Noterdaeme+11}.

At lower redshift we use the measurements from the SZ effect.
Thus, spectral measurements of galaxy clusters at
different frequency bands yield independent intensity ratios for
each cluster. The combinations of these measured ratios permit to
extract the cosmic microwave background radiation (see Fabbri et
al. \cite{fab78}). We relied on the data compilation in Luzzi
et al. \cite{luz09}, which have analyzed the results of
multifrequency SZ measurements toward several clusters from 5
telescopes (BIMA, OVRO, SUZI II, SCUBA and MITO).

\begin{figure}[t]
\vspace{0.5cm} \centering
\includegraphics[scale=0.6,angle=0]{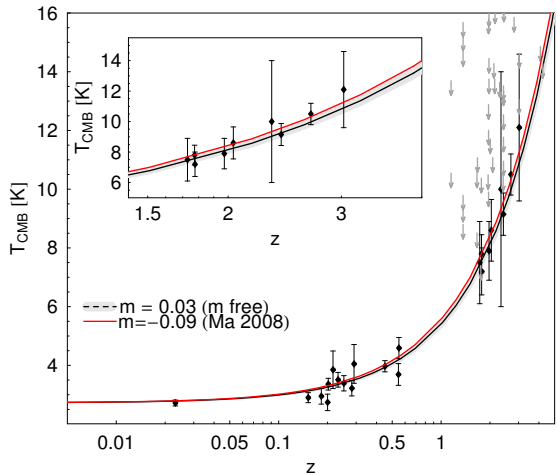}
\caption{Cosmic microwave background temperature as a function of
the redshift. The black points with bars are the full collection
of measurements from Luzzi et al. \cite{luz09} and Noterdaeme et
al. \cite{Noterdaeme+11}. The gray arrows represent the upper
limits derived from the analysis of atomic carbon (see
\cite{Noterdaeme+11} for details). The black line is the best fit
result ($m=0.03$), while the gray region is the $1\sigma$
uncertainty. The red line is the best fit recovered from Ma
\cite{ma} (see the legend). The inset panel show a magnified
vision of the higher redshift region of the plot.}
\label{fig:figT}
\end{figure}

We derive the observed $T_{CMB}$ from the theoretical
expression $T_{th}$, which we have derived in Eq. (\ref{ma3}).
We set $T_0 \, = \, 2.725 \ {\rm K}$, which
is quite well determined in the literature \cite{mat99}, and the
matter density $\Om=0.273$ to the value inferred in Komatsu et al.
\cite{komatsu+09}. If we take $\gamma=4/3$, then we find $m =
0.03^{+0.08}_{-0.09}$ \cite{Jetzer}, which is lower than the estimated value of
$m=0.09 \pm 0.10$ in \cite{Jetzer+11}, but fully consistent within
uncertainties, and also pretty consistent with $m = 0$. In Fig.
\ref{fig:figT} the temperature measurements (together with some
upper limits) are shown, and our best fitted result is plotted and
compared with the $m=-0.09$ result in Ma \cite{ma}. The value we
have found corresponds to an effective equation of state $\weff =
-0.99 \pm 0.03$, consistent with $\weff = -1$, and the transition
redshift is $z_{T} = 0.78 \pm 0.08$. In order to check the impact
of redshift distribution we separate the data in two redshift bins
with $z < 0.6$ (SZ data only) and $z \geq 0.6$ (QSO absorption
lines only), finding the best fitted values
$m=0.12^{+0.12}_{-0.13}$ and $-0.05^{+0.12}_{-0.14}$,
respectively. Although the uncertainties are very high, it seems
to emerge a mild trend with lower redshift data preferring a DE
decaying into matter and radiation, while data at $z > 0.6$ point
to an opposite behavior. These results could be interpreted in a
different way, in fact, the differences found could be due not to
the different redshift coverage, but to some particular biases in
the two kind of observations, SZ vs QSO absorption lines. An
indication of this suggestion comes if we divide the sample in
three subsamples: 1) SZ data, 2) the data from the analysis of the
fine structure of atomic carbon (AC), and 3) the measurements
based on the rotational excitation of CO molecules in
\cite{Noterdaeme+11}. If we fit the model to the combined SZ+AC
and SZ+CO samples we find $m = 0.11_{-0.11}^{+0.11}$ and $m=
0.03_{-0.10}^{+0.09}$, respectively. Because of the larger
measurement errors, the AC data affect very little the results
when only SZ measurements are adopted, and the result for the
second sample shows that SZ and CO data mainly constrain $m$, and
AC simply gives a tiny reduction on the errors.

Adopting a constant value for the ratio $\psi_{\gamma}/3n_{\gamma}H=\beta$
Lima et al. \cite{lima+00} have found the simple relation,
\begin{equation}
T(z)=T_0 (1+z)^{1-\beta}~. \label{d1}
\end{equation}
If we fit all the sample we find $\beta = -0.002 \pm 0.03$, while
for the low and high redshift subsamples we have $\beta = 0.06 \pm
0.08$ and $-0.01 \pm 0.03$, respectively. These results are
qualitatively consistent with what found in Noterdaeme et al.
\cite{Noterdaeme+11}, and points to a similar trend as the one
discussed above.

If $\gamma$ is left free to change, for the best
fitted value it turns out that $\gamma > 4/3$ and $m$ is systematically more positive.
We find
$\gamma = 1.35_{-0.03}^{+0.03}$ and $m= 0.25_{-0.17}^{+0.23}$,
which corresponds to an effective equation of state $\weff = -0.92
\pm 0.07$ and the transition redshift is $z_{T} = 1.1 \pm 0.6$.
When the two subsamples are adopted, wide confidence contours are
found, particularly for the $z \geq 0.6$ sample, for which the
contours at very low $m$ are not closed.  We obtain $\gamma =
1.3_{-0.1}^{+0.2}$ and $1.26_{-0.01}^{+0.01}$, while $m =
0.8_{-0.3}^{+0.1}$ and $0.6_{-1.0}^{+0.1}$, respectively for the
low and high-z samples.

We notice that if DE does not decay into radiation (corresponding
to $\epsilon=\tilde{\epsilon}=0$ and thus $m$ is no longer
constrained) then the CMB temperature will scale in the standard
way. Clearly, this would imply that if DE decays, this has to be
into DM only. On the other hand a deviation  of the CMB
temperature from the standard scaling could be interpreted as DE
decaying also into radiation. In which case with Eq. (\ref{ma3})
one can determine either $m$ or $\tilde{\epsilon}$ and thus get
some insights on the decay mode of DE into radiation. Future data
on the CMB temperature will allow to shed light on this important
issue.\\
In \cite{Jetzer} we have combined the CMB temperature data with a set
of other independent measurements such as Supernovae Ia distance moduli,
CMB anysotropy, baryon acoustic oscillation and observational data
for the Hubble parameter. When combining all
the data sets we found values for $m$, and thus $\weff$
consistent with the ones obtained when using only CMB temperature data.

C. Tortora was supported by the Swiss National Science Foundation.\\

\end{document}